\documentclass[english]{article}
\usepackage[T1]{fontenc}
\usepackage[latin9]{inputenc}
\usepackage{array}
\usepackage{multirow}
\usepackage{amsmath}
\usepackage{amssymb}

\makeatletter

\providecommand{\tabularnewline}{\\}

\makeatother

\usepackage{babel}
\usepackage[a4paper, margin=25mm]{geometry}
\begin{document}
\emergencystretch=1.5em

\title{Strategy for the Search for Communicative Benevolent Extraterrestrial
Intelligence for Mitigation of Anthropogenic Risks}

\author{Aleksey Borisenko}
\maketitle
\begin{abstract}
Despite seven decades of dedicated SETI efforts, observational data
regarding the presence of communicative extraterrestrial intelligence
(ETI) within our immediate cosmic neighborhood remains inconclusive.
A fundamental vulnerability of contemporary SETI paradigms lies in
the absence of conclusive search strategies backed by formal logic.
Ways of reasoning for such strategies are typically obscured when
attempting to search for all conceptually possible variations of ETI
simultaneously.

This paper demonstrates that by limiting the target profile to a specifically
constrained class of benevolent communicative ETI, several critical
conclusions can be derived: (1) an effective and economically viable
SETI strategy becomes achievable; (2) the probability of such an ETI
existing within a 100-light-year radius approaches unity; (3) this
subclass of ETI represents the optimal and secure target for contact;
(4) the execution of this contact strategy serves as a cost-effective
preventive measure against anthropogenic existential risks and accidental
interstellar conflicts; (5) SETI concepts currently deployed have
a negligible probability of detecting the anticipated beacon signal,
as it is expected to be either intermittent or below their noise floor.

Incapacity of conventional SETI methods is explained by conservative
adherence to historical concepts without enough critical analysis
of them from the perspective of contemporary state of technological
trends.
\end{abstract}

\section{Introduction}

The inception of SETI in the mid-20th century was driven by the realization
of humanity's technical capacity to establish interstellar radio communication
links. However, at that historical juncture, humanity remained far
from initiating any extrasolar colonization effort. Consequently,
the only immediate application of interstellar telemetry was the search
for conjectured ETI.

While initial public and academic sentiment was characterized by optimism
regarding the technological and societal benefits of contact, this
enthusiasm was subsequently tempered by skeptical counterarguments.
This shift led to the prevailing paradigm of skepticism that currently
dominates political, scientific, and public discourses. As a consequence,
active SETI research has been marginalized into a few narrow niches,
surviving primarily due to its relationship with general observational
astronomy, science fiction, and popular science platforms. Frequently,
these domains utilize the public appeal of SETI as a mechanism to
attract funding, without pursuing the core operational goals of SETI
or exploring specialized methodologies from which symbiotic domains
derive no utility. Under such conditions it can be reasonably suspected
that incapacity of contemporary SETI may be explained by metascientific
problems.

One revealed class of metascientific problems is using historically
established logical concepts proposed by pioneers of SETI in mid-20th
century without enough critical analysis of their grounds according
to evident contemporary trends. Just for example, let's look at the
Dyson Sphere concept here. Given the well-known problem of technogenic
orbital debris around Earth, it seems reasonable to suspect that extrapolation
of such a problem to the scale of a Dyson Sphere will show that Dyson
Spheres are physically impossible; thus it is the responsibility of
any contemporary work relying on a Dyson Sphere to show feasibility
of this concept, not just to refer to prior works where the same concept
was used. Instead, there is a well-established practice to ignore
this intuitively visible problem, resulting in a situation where nobody
tried to publish an appropriate critical audit of the Dyson Sphere
concept, because none of the funding symbionts of SETI are interested
in such an audit. A similar situation is evident with many historical
speculative concepts of the 20th century, most notable of which are
Bracewell Probes, Von Neumann Probes, and the Kardashev Scale.

Meantime, design vectors of contemporary SETI projects are still defined
by assumptions drawn from such unreliable historical concepts. Moreover,
\emph{the well-established dominating point about the absence of practically
contactable benevolent and trustable ETI in our proximity is logically
grounded on those unreliable concepts, making our civilization blind
to possibilities which such ETI can provide for us for mitigation
of our anthropogenic existential risks}.

This paper pursues a constructive approach for solving revealed metascientific
problems by building independent formal logic from the ground up to
avoid relations to unreliable historical concepts. The resulting novel
logical framework is introduced in Section 2 along with a brief explanation
of key points of failure of the traditional approach.

The proposed structure of research consists of the following steps:
(1) engineering of an optimal contact strategy based on available
technological knowledge; (2) examination of evolutionary mechanisms
leading to probable existence of an appropriate class of ETI in our
proximity; (3) examination of motivations of both ETI and emerging
self-evolved civilizations to support that contact strategy; (4) examination
of compatibility of the resulting hypothesis with known empirical
facts; and (5) attempts of observations directed by known results
of engineering. Those steps are organized as an engineering cycle.
Perhaps this metacognitive structure of an engineering cycle is not
invented here, but rather explicitly formalized. In general, this
engineering cycle somehow works without concise formalization, but
this formalization may help revealing and solving metacognitive problems
of SETI. It appears that the cycle is now almost halted due to systematically
frustrating results of examinations based on all known results of
engineering. And now some persistent problems can be formally outlined.

In the 20th century, we were for a long time in a situation of fast,
apparently perpetual advance of technological knowledge, where it
was logically impossible to infer any stably optimal contact strategy
without speculations about continuously operating beacons based on
far superior technological knowledge of ETI, which is not yet understood
by our civilization. In this situation, the only feasible effective
strategy of SETI relied on the capacity of ETI to predict general
astronomical instrumentation developments of emerging young civilizations
and target the most optimal of those instruments. But examination
of this strategy at steps 2, 3, and 4 gives frustrating results, leading
to demotivation of the whole engineering cycle, while conservative
support of a suspicious historical SETI strategy gives some additional
public feedback for general astronomy for negligible cost. Finally,
when flaws of the historical SETI strategy appeared too obvious, general
astronomy switched its attention from intentionally supported beacons
to unintentional technosignatures.

But now technological knowledge appears stable enough for a decade;
therefore resuming the engineering cycle now can give an affordable
SETI strategy which can be proven effective. The current situation
allows two enabling possibilities: (1) use of intermittent beacons,
dramatically reducing the cost of beacon support for ETI; and (2)
deduction of many parameters of the beacon signal from optimality
criteria based on stable technological considerations of a young civilization,
dramatically reducing the search space for SETI.

\section{Decomposition of the Art}

\subsection{Resolution of the Fermi Paradox}

The historical debate surrounding the apparent absence of ETI is encapsulated
by the Fermi Paradox, initially framed by Michael Hart\cite{Hart (1975)}
as ``Fact A'' --- the observation that extraterrestrials are not
physically present on Earth. Hart categorized potential explanations
into technical, sociological, observational, and temporal. Then Hart
briefly analyzed all of them, suggesting none of them satisfactory;
thus the most probable cause of Fact A is the absence of ETI in the
Galaxy. Hart recognized the brief nature of his analysis; hence his
conclusion remained questionable. Then Tipler\cite{Tipler (1980)},
based on speculation about self-replicating probes, concluded with
greater confidence that space-faring extraterrestrials do not exist
in the Galaxy. From this point, SETI enthusiasts faced a serious problem
with argumentation of their efforts. They were forced to search for
weak points in the logical framework defined by Hart and Tipler, and
were typically limited to two simplest options: (1) interstellar expansion
is not possible; or (2) using a slow expansion model, argued by extreme
technical difficulty of expansion, thus the temporal explanation becomes
a feasible basis for SETI. With either of those options taken, we
must conclude that the chance of having any ETI in our close proximity
will be marginally small; thus we must look as far as possible through
the whole Galaxy to improve our chances. Yet one known option to explain
``Fact A'' for SETI is based on simulations proving the possibility
of occasionally unsettled systems, but those simulations, like Carroll-Nellenback
et al.\cite{Carroll (2020)}, use arbitrary mechanistic expansion
models which appear unrealistic from an economical and motivational
perspective of intelligent civilization. After substantial SETI efforts,
Hart's logical framework was generalized by using the fact of the
``Great Silence'' instead of ``Fact A'', accounting for the absence
of detected signals from ETI.

But in general, when hypothesizing a dominant galactic-wide culture
supported by ETI, it is logically not necessary to adhere to the specific
logical framework with categorizations proposed by Hart. Instead,
the proposed cultural framework governing interstellar transactions
must be shown to (1) possess a competitive advantage, allowing it
to naturally evolve into the dominant paradigm across the cosmos;
and (2) explain ``Fact A'' and the ``Great Silence''.

Basic prerequisites of a competitive interstellar culture are (1)
mitigation of existential risks and (2) economical reasoning and optimization
of expensive operations.

The competitive advantage of a dense network of benevolent communicative
ETI is well shown by Korhonen\cite{Korhonen (2013)} as following
from their unique collective defense capability against pathologically
hostile competitors. Self-replicating probes used by Tipler in his
reasoning fall in the category of such pathologically hostile competitors;
they will be destroyed by the dominant culture of ETI.

Expansion of ETI may be motivated by achievement of the following
goals:
\begin{enumerate}
\item Mitigation of planetary existential risks by diversification of living
environments and planetary cultures for which it is problematic to
coexist on a single planet.
\item Extending the basis for mining empirical data about existential risks.
\item Reducing the expense of interstellar communication traffic by shortening
communication links.
\item Reducing the fraction of nearby unpopulated planets where hostile
civilizations may occasionally evolve to an optimally manageable level.
\end{enumerate}
When ETI density reaches a certain level, somewhere between 1\% and
99\% of habitable planets, expansion will be stopped due to the absence
of economical return, explaining Fact A. It can be shown that the
expansion process will fill the whole habitable zone of the galactic
disk to that saturated state during a billion years. Even if such
spatial expansion is not a goal for ETI, it will be achieved as a
side effect when ETI tries to keep optimal local population density
on the edge of the populated zone, due to natural random motion of
populated stars. This means that presence of ETI within 100 ly is
highly probable, and there is no sense to look beyond that range,
and we have only 1000 candidate stars to look for a signal from ETI.

\subsection{The Bottom-Up Approach}

To date, nearly all empirical SETI initiatives have based their search
methodologies on a bottom-up framework. This paradigm initiates from
an exploration of humanity's current technological capabilities to
capture cosmic signals, under the assumption that an ETI will choose
a signaling mode compatible with the technical apparatus of a young
civilization. Within this baseline framework, the technical capabilities
of an ETI are assumed to be far superior to our own, rendering specific
engineering details unpredictable. Consequently, our search strategy
is guided primarily by technical limitations of existing observation
tools optimized for current needs of general astronomy. Due to the
intermittent nature of astronomical observations, ETI must indefinitely
support expensive dedicated beacon hardware for each target to comply
with our bottom-up approach; therefore it seems intuitive that this
strategy is likely suboptimal for ETI and therefore not used.

Instead, ETI likely will use an optimal contact strategy intended
to contact technologically mature young civilizations only, where
technological capabilities of both sides are already well bounded
by common laws of physics; thus both sides can converge to the same
economically optimal contact strategy.

Here it is worth clarifying the meaning of the term ``technological
level'' of a civilization in this context. Here technological level
is shaped not by capabilities of technological goods already physically
built, but by capability to assess what and how can be built economically
given enough time and resources at the current technological level.
Note that in this sense our civilization already appears on the verge
of maturation for a decade, as scaling rules and limits of most demanded
technological possibilities are already well investigated and it is
well known how they are shaped by well-known laws of physics. Moreover,
a symptom of our maturation is clearly visible as a trend of slowing
down of many technological benchmark indices; they are already an
order of magnitude slower than 30 years ago. Thus, it seems appropriate
now to use an appropriate approach to SETI.

\subsection{The Top-Down Approach}

This paper introduces a top-down approach to SETI, beginning with
logical deductions regarding the nature and operational constraints
of an ETI. From these foundational premises, an optimal interstellar
contact strategy can be derived --- one that an ETI would deploy to
communicate specifically with technologically mature young civilizations.
The top-down framework assumes humanity can reach this state of technological
maturity within the next decade, given an affordable engineering effort.

Critics may argue that our apparent trend of maturation may be just
a transition into a metastable state, where further transitions are
still possible, say after 100 or 1000 years, due to invention of new
technological paradigms; thus supertechnological ETI still may use
exotic communication methods beyond our current paradigm which are
not yet discovered by a young civilization. However, this objection
ignores a fundamental strategic possibility: such supertechnological
ETI has the choice to comply with an effective top-down strategy which
can be inferred by a typical young civilization in a metastable state
because that young civilization principally cannot distinguish this
state from absolute maturation. Thus, the top-down approach will be
the optimal contact strategy for a young civilization on any apparently
metastable (quasimetastable) technological level.

Concurrently, anthropogenic existential risks of young civilizations
in a quasimetastable state may be acute. Trying to contact absolutely
mature civilizations only, ETI may reduce chances of contact due to
those anthropogenic risks. Thus, especially if those risks are objectively
acute, ETI likely will try to contact young civilizations timely on
the verge of a typical quasimetastable state of a young civilization
if it is affordable; otherwise a young civilization may irreversibly
degrade into a mundane non-contactable, or worse, into a pathologically
hostile state.

Degradation of an uncontacted young civilization during a quasimetastable
state is reasonably expectable. Diminishing economical return from
innovative activity will naturally strip investments into non-mundane
projects if they are not protected by explicit policy. Concurrently,
advocates of non-mundane projects may escalate strong social polarization
and severe instability until benefits of non-mundane existence are
clarified by trustable empirical proof. Support of a logically proven
optimal contact strategy may help, but some civilizations may fail
to support it for long, either irreversibly falling into mundane existence,
or concluding loneliness or the Dark Forest hypothesis. Using an optimal
contact strategy, timely detection of ETI can prove marginality of
Dark Forest and Black Ball hypotheses\cite{Bostrom (2019)} relaxing
fatalistic fears and unnecessarily strict anti-technological policies.
Then ETI can give optimal cures against possible dark balls, trying
to align interstellar behavior of a young civilization with established
ethical norms of the communicative population of the Galaxy.

\subsection{Expected Nature of ETI}

Any civilization that develops space-faring capabilities inevitably
achieves the physical capacity for self-destruction through interstellar
warfare against its own former colonies escaped from hegemonic control
of the parent planet. Once started, interstellar expansion goes out
of hegemonic control of the starting planet almost immediately, and
any civilization will necessarily recognize this at the preliminary
stage of development of the first interstellar expansion mission.
Then the starting planet will try to develop some realistic non-hegemonic
model of interstellar cooperation which is still worthy of interstellar
expansion missions. It will be recognized that unavoidable uncontrolled
runaway expansion will establish a condition where dominant galactic
intelligent species will evolve genetically to be more capable for
interstellar expansion than species remaining at the starting planet,
making the starting planet noncompetitive in peer-to-peer warfare
against a typical planetary civilization. Then the starting planet
can survive only in conformity with the naturally evolved dominant
cooperative interstellar culture, due to the unique collective defense
capacity of that culture.

Given a substantial fraction of hostile space-faring civilizations
still unavoidably emerging, any civilization has little chance to
survive for long without stable cooperative relations with a few nearby
planets to take competitive advantage of collective defense. Even
assuming an evolution model of non-intelligent civilizations under
such conditions, where non-hostile cooperative behavior appears as
a randomly emerging gene, this gene will have a competitive advantage
and species with this gene will be dominant. Reasonably, any planetary
civilization must stay intelligent for a long time mastering appropriate
technologies to be physically capable to be hostile to other stellar
civilizations; thus it must care about its own longevity. Thus, it
must recognize the value of that ``gene'' of non-hostile cooperative
behavior, and it will expose it in any interstellar action. Thus,
hostile behavior is possible only as a rare statistical anomaly. As
shown by Korhonen\cite{Korhonen (2013)}, there is no need for any
arms on duty to support interstellar collective defense protocols.
Any specific arms-building infrastructure can be built for retaliatory
response only on demand when a nearby pathologically hostile actor
is revealed. The resulting population of ETI will be dominated by
benevolent planetary civilizations interconnected into a galactic
cooperative network. Dominant ETI will expose predictable, trustable
behavior which will be investigated further in this article, while
the rest of the population is composed of rare highly unpredictable
outsiders.

Two kinds of rare galactic outsider can be hypothesized based on known
SETI-related arts. One expectable mode of anomaly is a young space-faring
civilization pathologically adhered to the culture of Dark Forest,
making a preemptive attack against an occasionally detected beacon-supporting
civilization. A second example is a young civilization erratically
concluding loneliness in the Galaxy and starting careless expansion.
Both cases can be qualified as existential failures of a young civilization.

A common counterargument, famously popularized by Stephen Hawking,
posits that attempting contact with an ETI presents a severe existential
threat to humanity, citing historical analogies of terrestrial colonization.
However, such arguments lack comprehensive formal validation. The
underlying logic relies on a primitive extrapolation of human geopolitical
behavior onto an advanced extrasolar civilization. While competitive
and hostile behaviors are prevalent among terrestrial species, they
are a suboptimal byproduct of survival strategies driven by resource
scarcity. Hostility leads to an escalation of conflicts that is mutually
destructive to all participating parties. Therefore, any sufficiently
mature intelligent civilization will naturally seek to mitigate conflict
vectors where possible. Interstellar hegemony or material extraction
is highly improbable to yield a net positive economic return for a
planetary civilization, due to the extreme expenditure of interstellar
transport. Since there are no shared, physically accessible expendable
resources to spark a direct conflict, traditional terrestrial drivers
of aggressive expansion do not apply.

Consequently, a benevolent, communicative ETI will prioritize the
establishment of a dense, cooperative galactic network, which offers
a collective defense advantage against any anomalous, pathologically
hostile civilizations. Maintaining strictly honest ethical standards
within interstellar transactions is functionally mandatory for such
a cooperative network to ensure trust, especially since physical verification
of received telemetric data is impossible.

\subsection{Contact Strategy}

An interstellar communication transceiver facility appears mandatory
for continuous support by each planetary civilization of communicative
ETI. If it is useful sometimes as a beacon to support contact strategy,
it appears as the most economic kind of beacon. From there it can
be deduced that the narrow 1030 nm line of a Yb:YAG laser will be
used in the beacon, because it can be argued as an optimal kind of
laser for interstellar communication\cite{Hecht(1996)}. Such a beacon
must be rare and intermittent, where periodicity on the order of a
decade and duration on the order of 24 hr can be argued. Then, the
``Great Silence'' will be explained by a lack of appropriate SETI
efforts to detect this specific kind of signal. Even if a signal is
occasionally detected, immediate follow-up observation is required
by a special large message-receiving telescope for qualification of
that signal; otherwise it will be reasonably qualified as a glitch
of unknown nature, as already happened with certain Wow! signals.
That signal must be followed by a message with technical instructions
about making a response, and, maybe, some kind of urgent data typically
useful by young civilizations for mitigation of their existential
risks. As an example, this may be statistics of typical failure modes
of young civilizations within a century of contact, though it may
be problematic to explain failure modes in a small message.

\section{Interstellar Communication System Engineering}

For interstellar communications, a planetary civilization will support
an appropriate transceiver at Lagrange point L2, composed of a large
segmented-mirror telescope and some number of Yb:YAG laser beamlines
whose outputs are spectrally combined in a Dense Wavelength Division
Multiplexing (DWDM) configuration. Let $A$ be the telescope area
and $N$ be the number of channels. Mean power of a single beamline
is fixed to $P=2\text{ kW}$ as a realistic engineering baseline grounded
in demonstrated cryogenic helium-gas-cooled multi-slab Yb:YAG amplifier
architectures (such as the DiPOLE class)\cite{Mason (2017),Banerjee (2020)},
which achieve kilowatt-class average power with high optical-to-optical
efficiency while controlling amplified spontaneous emission (ASE),
thermal lensing, and stress-induced birefringence\cite{David (2018)}.
Scaling a single high-energy pulsed beamline far above the multi-kilowatt
level degrades beam quality and increases catastrophic failure risk,
whereas parallelizing multiple $2\text{ kW}$ multi-slab beamlines
via DWDM preserves near-diffraction-limited focusability and modular
redundancy.

\subsection{Optical Link Analysis}

Let $p$ and $n$ be photons per second of signal and noise respectively
detected by one receiver channel, and $t$ be slot time; then Photonic
Information Efficiency (see Appendix \ref{sec:Appendix-A}) can be
expressed as 

\begin{equation}
\Xi=1.22\left(-\ln(3nt+pt)-1\right)=1.22\left(-\ln(3n+p)-\ln(\mathrm{e}t)\right)
\end{equation}
and bitrate can be calculated as

\begin{equation}
R=pN\Xi=1.22pN\left(-\ln(3n+p)-\ln(\mathrm{e}t)\right)
\end{equation}

\subsection{Starlight Photon Rate}

Let an ideal antenna of area $A$ be pointed to a star at distance
$d$ with projected area of star $A_{s}=\pi R_{s}^{2}$ and temperature
$T_{s}$. Then the received solar photon rate at single-mode output,
in both polarization modes, in bandwidth $B$, can be expressed as

\begin{equation}
n=\frac{2A_{s}}{\exp\left(\frac{hc}{\lambda kT_{s}}\right)-1}\frac{BA\eta_{r}}{\lambda^{2}d^{2}}=Bs\frac{A\eta_{r}}{d^{2}}
\end{equation}
where $k$ is the Boltzmann constant and $s=\frac{2A_{s}}{\exp\left(\frac{hc}{\lambda kT_{s}}\right)-1}\frac{1}{\lambda^{2}}$
is the absolute spectral brightness of the star, with typical value
$s\approx10^{29}$ for a Sun-like star at 1030 nm.

When the telescope is pointed on a planet separated from its host
star by projected distance $r$, spatial resolution against starlight
can be defined by the resolution parameter

\begin{equation}
x=\alpha\frac{2\pi R}{\lambda}=\frac{2r\sqrt{\pi A}}{\lambda d}
\end{equation}
and then theoretical sensitivity of an ideal round telescope to starlight
is defined by the corresponding PSF: $\left(2\mathrm{J}_{1}(x)/x\right)^{2}$,
but for a real telescope it is practical to use a monotonic envelope
of that PSF instead, which can be approximated as:

\begin{equation}
E(x)=\begin{cases}
1-(0.4x)^{2} & \text{if }x<2\\
2.545/x^{3} & \text{otherwise}
\end{cases}\approx\begin{cases}
1 & \text{if }x<1.36\\
2.545/x^{3} & \text{otherwise}
\end{cases}
\end{equation}
and then the suppressed stellar photon rate can be estimated as:

\begin{equation}
n=E\left(\frac{2r\sqrt{\pi A}}{\lambda d}\right)Bs\eta_{r}\frac{A}{d^{2}}
\end{equation}
and for the spatially resolved case ($x>1.36$) 

\begin{equation}
n=\frac{2.545}{x^{3}}\frac{B}{N}s\frac{A\eta_{r}}{d^{2}}=\frac{Bs\lambda^{3}d\eta_{r}}{17.5r^{3}N\sqrt{A}}
\end{equation}
where $B=113\cdot10^{9}\text{ Hz}$ is the useful bandwidth of the
Yb:YAG lasing medium, occupied by $N$ DWDM channels; thus single-channel
bandwidth is $B/N$. Then assuming number of channels $N$, bitrate
is
\begin{equation}
\mathrm{R}(A,d,N)=\begin{cases}
p=\frac{P\eta_{c}A^{2}}{hc\lambda d^{2}}\\
n=\frac{Bs\lambda^{3}d\eta_{r}}{17.5r^{3}N\sqrt{A}}\\
1.22pN\left(-\ln(3n+p)-\ln(\mathrm{e}t)\right)
\end{cases}
\end{equation}

\subsection{Starlight Factor}

For the spatially resolved case, the influence of residual diffracted
starlight can be expressed by the starlight factor

\begin{equation}
S=\frac{3n}{p}=\frac{3Bs\lambda^{4}d^{3}\eta_{r}hc}{17.5r^{3}PN\eta_{c}A^{2.5}}=200\frac{L^{3}}{ND^{5}}
\end{equation}

If $S<1$, then $3n<p$ and starlight can be neglected, while for
$S>1$, PIE depends mostly on starlight.

As an initial coarse guess, let $L=50\text{ ly}$, $D=20\text{ m}$
and $N=16$ because it is a reasonably expectable case, which can
be argued by principles of the top-down approach, economical model
of expansion, and engineering intuition. Then $S_{r}=0.49$ appears
at the edge of two regimes. For smaller systems PIE will be defined
solely by starlight, as $\Xi=-1.22\ln(3\mathrm{e}nt)$, while for
larger systems it is independent of starlight, $\Xi=-1.22\ln(\mathrm{e}pt)$.
This edge appears as an economical counterargument against downscaling
of system size beyond this edge, because downscaling rules become
worse there. Thus, $S\approx1$ may appear useful as an additional
Schelling point of contact strategy, for inference of a lower bound
on beacon size. With the $S\approx1$ assumption, PIE can be expressed
as $\Xi=-1.22\ln(2\mathrm{e}pt)\approx15$, and bitrate can be estimated
as $R=pN\Xi=207{,}000$ bits per second. Note that PIE is a complicated
but slow logarithmic function in the range of our interest, peaking
at $S\approx1$; therefore for the purpose of coarse estimations,
PIE can be assumed constant, say $\Xi=10$, when used for bitrate
estimation by the $R=pN\Xi$ relation in a rather wide range of conditions.
For instance, assuming $\Xi=\text{const}$, the simple $A=2KN$ relation
can be proven as a near-optimal economical solution in the next subsection.

\subsection{Optimization of Telescope Area}

Let the cost of maintenance per square meter of telescope be 1, and
the cost of maintenance of one laser beamline be $K$. Total cost
is $C=KN+A$. For a given bitrate ($R=\text{const}$), $G=NA^{2}=\text{const}$,
$N=G/A^{2}$,

\begin{equation}
C=\frac{KG}{A^{2}}+A
\end{equation}

\begin{equation}
\frac{dC}{dA}=1-\frac{2KG}{A^{3}}
\end{equation}
Then optimal $A$ can be found from the solution of $\frac{dC}{dA}=0$:
\begin{equation}
\begin{cases}
A=G^{1/3}(2K)^{1/3}\\
N=G^{1/3}(2K)^{-2/3}
\end{cases}
\end{equation}

\begin{equation}
A=2KN
\end{equation}
Thus, the cost of the telescope is double the cost of lasers, $C=3KN=1.5A$.

Formally, the $A=2KN$ solution is valid for a discrete set of given
bitrate values, where $A=2KN$ gives a natural value of $N\in\mathbb{N}$.
But a straightforward practical engineering solution is rounding to
the nearest bitrate from the available set, and this solution remains
near-optimal if $N>1$ and asymptotically optimal for $N\gg1$. Then,
$A\approx2KN$ must be valid approximately for an optimal system with
$N>1$, and therefore this relation is useful for estimation of beacon
parameters of ETI.

\subsection{Laser Cost Estimation}

Cost of long-term maintenance of a large space-based facility can
be estimated from mass flow of spare parts and materials needed for
maintenance of such a facility.

Key spare parts of a large telescope are segments of the primary mirror,
with a large delicate optical surface, say $1\text{ m}^{2}$ per segment,
exposed to micrometeoritic damage, and therefore they need replacement,
say once per decade.

Lasers have no such large delicate optical surface exposure, but they
have large exposed radiators which can be dramatically damaged by
rare meteoritic hits, say once per 1000 years per square meter, with
loss of the whole charge of coolant. Segmentation of radiators is
very problematic because each segment needs a hermetic tube connection.
Replacement may be practical for the whole radiator of a contour,
whose cost is comparable to the cost of the whole beamline. Moreover,
each cooling contour requires a mechanical pump with highly loaded
bearings exposed to pure helium coolant. Use of any lubricants is
problematic in such bearings due to contamination of coolant, with
following contamination of optical surfaces. There are two options:
levitation-type bearings or lubricant-free mechanical bearings. Maintenance
of the laser system may be composed of three distinct parts:
\begin{itemize}
\item Replacement of the whole beamline on severe meteoritic damage.
\item Regular replacement of bearings, say every year, though bearings are
small parts and their mass flow is little.
\item Regular supply of fresh helium required for any on-board purification
system.
\end{itemize}
Perhaps each part is worthy of distinct deep research, but here I
limit to rough estimation, by intuitive guess that maintenance cost
of the laser is dominated by the first part, where beamline lifetime
per square meter of radiator area is $\Lambda_{L}=1000\text{ yr}\cdot\text{m}^{2}$,
lifetime of primary segments is $\Lambda_{t}=10\text{ yr}$, and solar
panels $\Lambda_{p}=30\text{ yr}$.

\subsubsection{Size of Radiators}

Radiation power per square meter of a two-sided black-body radiator
is $2\sigma T^{4}$, where $\sigma=5.67\cdot10^{-8}\text{ W m}^{-2}\text{ K}^{-4}$
is the Stefan--Boltzmann constant. Yb:YAG laser amplifiers are much
more efficient at $T\approx100\text{ K}$ than at room temperature,
but due to the cost of high-power cryogenics, the economically optimal
temperature is usually considered to be around $T\approx150\text{ K}$.
Only a small fraction of heat, about $f_{c}=0.25$, is produced by
the crystal, while the rest is produced in pump diodes, operating
near room temperature. Diodes likely will be fiber-coupled and distributed
along the surface of a dedicated radiator, to avoid problems of a
liquid-cooled system in space (terrestrial laser facilities normally
use water-cooled systems instead). State-of-the-art high-power Yb:YAG
laser amplifiers normally use helium gas to cool crystals, and this
approach seems well suitable for a space-based laser. Assuming a realistic
temperature drop in both radiators of 20 K, temperatures of cold and
hot radiators will be $T_{c}=130\text{ K}$ and $T_{h}=280\text{ K}$
respectively. Then radiator area is

\begin{equation}
A_{r}=\frac{P_{b}}{2\sigma}\left(\frac{f_{c}}{T_{c}^{4}}+\frac{1-f_{c}}{T_{h}^{4}}\right)\left(\frac{1}{\eta_{L}}-1\right)=70\text{ m}^{2}
\end{equation}
where $P_{b}=2000\text{ W}$ is the power of a single beamline. Solar
panel area per beamline is

\begin{equation}
A_{p}=\frac{P_{b}}{I\eta_{L}\eta_{p}}=50\text{ m}^{2}
\end{equation}
where $I=1000\text{ W/m}^{2}$ is typical insolation in the habitable
zone, and$\eta_{p}=0.2$ is efficiency of the solar panel. Then cost
of the laser (relative to a square meter of primary segments) is

\begin{equation}
K=\Lambda_{t}\left(\frac{A_{r}^{2}m_{r}}{\Lambda_{L}}+\frac{A_{p}m_{p}}{\Lambda_{p}}\right)\approx10\text{ m}^{2}
\end{equation}
where $m_{r}=0.2$ and $m_{p}=0.1$ are specific mass factors of radiator
and solar panel respectively, $\Lambda_{t}=10\text{ yr}$ and $\Lambda_{p}=30\text{ yr}$
are lifetimes of primary and solar panel segments respectively, and
$\Lambda_{L}=1000\text{ yr}\cdot\text{m}^{2}$ is specific lifetime
of the laser per square meter of radiators. Cost of the solar panel
appears negligible here, because it is easily segmentable and lightweight.

From $A=2KN$ it follows that total radiator area of lasers, $A_{r}N=\frac{A_{r}}{2K}A$,
is $3.5$ times larger than telescope area.

\subsection{Coarse Estimations}

Assuming $A=2KN$, $K=10$, $\Xi=10$ and $\eta_{c}=0.143$, bitrate
can be expressed as

\begin{equation}
\mathrm{R}=\Xi\frac{P\eta_{c}4K^{2}}{hc\lambda}\frac{N^{3}}{d^{2}}\approx62500\frac{N^{3}}{L^{2}}
\end{equation}
where $L$ is communication distance in light-years and

\begin{equation}
D=\sqrt{\frac{4A}{\pi}}\approx5\sqrt{N}
\end{equation}

Resulting bitrates for different values of $N$ and $L$ are presented
in the following table:

\begin{tabular}{|c|c|c|c|c|c|c|}
\hline 
 &  & N1 & N4 & N16 & N64 & N256\tabularnewline
\hline 
\hline 
Telescope diameter & $D=5\sqrt{N}$ & 5 m & 10 m & 20 m & 40 m & 80 m\tabularnewline
\hline 
Bitrate@3.2 ly & \multirow{4}{*}{$R=62500\frac{N^{3}}{L^{2}}$} & 6250 & 400k & 25.6M & 1.63G{*} & 105G{*}{*}\tabularnewline
\cline{1-1} \cline{3-7} 
Bitrate@10 ly &  & 625 & 40k & 2.56M & 163M & 10.5G{*}\tabularnewline
\cline{1-1} \cline{3-7} 
Bitrate@32 ly &  & 62.5{*} & 4k & 256k & 16.3M & 1.05G\tabularnewline
\cline{1-1} \cline{3-7} 
Bitrate@100 ly &  & 6.25{*}{*} & 400{*} & 25.6k & 1.63M & 105M\tabularnewline
\hline 
\end{tabular}

Note that 10 ly corresponds to a lower bound on average distance to
the nearest habitable exoplanet; 3.2 ly corresponds to rare close
encounters suitable for settlement; 32 ly corresponds to the most
expected distance to the nearest planet of ETI; and 100 ly corresponds
to the low-probability margin of our expectations from where ETI statistically
still may support a beacon for us.

Top-right cells with asterisks correspond to $\Xi\ll10$ where our
model is no longer accurate and where actual bitrate will be inferior.
Likewise, lower-left cells with asterisks may correspond to severe
pulse energy, exceeding 1 J, implying technical problems which are
not modeled here and will lead to inferior bitrate for a given cost.
Thus, there is a technological sweet valley along the backslash diagonal
of the table.

The problem of the top-right corner is not a high barrier for ETI
to increase bitrate if needed, and it is not a problem for SETI. That
problem is solvable by extension of bandwidth beyond the Yb:YAG line
by other slightly inferior kinds of lasers, such as Yb:YLF, Nd:YAG,
Yb:glass, or by second harmonics, etc., though Yb:YAG remains superior
among them, and size of the common telescope must be increased; therefore
power received by a SETI instrument within the narrow Yb:YAG band
will never decrease with increased bitrate of the communication system.
What is important for SETI is to know the lower bound on received
signal power within the narrow Yb:YAG band.

The problem of the lower-left corner may render smaller systems economically
infeasible, because their cost may appear almost independent of bitrate.
Thus, a reasonable lower bound on possible signal strength of a beacon
may be defined by the lower edge of the sweet valley. But this question
is not the subject of this article, because N16 can be reasoned here
as a lower bound on beacon-supporting ETI.

According to established principles of the top-down approach, a reasonable
lower bound N16 is argued well by physical size of the telescope comparable
to our expected next-generation flagship space-based telescope. Thus,
N16 appears close to our own communicativity on the verge of maturation.
N16 also famously appears in the middle of the technological sweet
valley, making sweet-valley assumptions appropriate for design of
a SETI mission.

\section{Economical Expansion Model}

Suppose we concluded that we are the only communicative intelligent
civilization in proximity (after appropriate SETI efforts) and likely
the only one in our Galaxy. Then it can be reasonably decided to make
a settlement each next 100,000 years to manage risks of our extinction.
We have a good chance to meet a suitable target within $3.2\text{ ly}$
during each 100,000 years.

A big interstellar ship will be assembled at Lagrange point L1 from
smaller parts conveniently launchable from Earth's surface. From L1
it will be slightly thrusted into interplanetary space, from where
it is already safe to use nuclear pulse propulsion in full thrust
mode.

Suppose we need 10,000 rocket launches to build an interstellar ship.
This corresponds to one rocket per decade on average, which seems
affordable. Suppose the risk-management problem is already solved
after a few such settlements and we have no more need for further
settlements for this particular purpose. But there may be a good opportunity
for doing some interstellar information business, using interstellar
communication.

Suppose N16 roughly corresponds to the cost of launching a rocket
once per decade and 25.6 Mbps at 3.2 ly. This seems affordable for
doing some interstellar business. But typical relative velocity of
a settled star is about 50 km/s. After the next million years, it
appears at distance 160 ly from us and bitrate drops to 10 kbps, meaning
most business opportunities are lost because cost of interstellar
traffic is no longer affordable for them. This business appears not
sustainable.

The average expense of settlements once per 100,000 years is one rocket
per decade, the same as support for N16; thus it is economically optimal
for interstellar communication business to keep interstellar distance
within $30\text{ ly}$ sustainably by means of regular settlements.
But such expense of settlements only exists on the edge of the already
populated zone; for a typical planet of ETI, it will cease after the
first settlement, forwarding the edge of the populated zone further
away from the planet with velocity comparable to average random velocity
of settled stars relative to the local standard of rest.

Such economically driven settlements will result in near-saturation
of the whole Galaxy after one billion years. Note that our model has
no implicit motivations of such spatial expansion. Each planet of
ETI just cares about optimal distance to a few closest neighbors,
but galactic-wide spatial expansion appears as a byproduct due to
natural random motion of stars, ``infecting'' the whole Galaxy with
this interstellar culture.

There is no economic driving force to settle every settleable planet,
and settlements must cease near saturation, when the fraction of populated
habitable planets (saturation level, $S$) appears somewhere between
$S=1\%$ and $S=99\%$. It already becomes hard to imagine economic
return from increasing local population density from 50\% to 90\%
if this requires interstellar settling expense.

ETI also can be motivated to leave enough room for self-evolving planetary
intelligence, because they can provide unique data of high value,
which cannot be mined by ETI in any other way. ETI will strongly adhere
to well-proven planetary sociopolitical cultures to minimize existential
risks, avoiding risky experimentations with them, while young uncontacted
intelligent civilizations are naturally forced to such risky experimentations
due to lack of appropriate empirical knowledge. But it seems nonsense
to leave more than 90\% of planets for this purpose. For the same
reason, ETI may be not interested in premature contact with young
civilizations, even if this is affordable for them; contact strategy
will be optimized to make contact at an optimized stage of evolution,
balancing between value of data mined by ETI per successful contact
and risk of existential failure of a young civilization to make appropriate
contact at all. Because success of contact strategy depends on the
technological level of a young civilization, the verge of technological
maturation is the last point for which technical aspects of strategy
can be optimized, and this is exactly the Top-Down strategy defined
in this article, by which we are now ready to contact.

The nearest planet of communicative ETI is expected to be most likely
within 30 ly, while 100 ly is a margin from where ETI still can support
a beacon for us due to rare statistical anomalies and our remaining
uncertainties. From this, a 50 ly range may be reasonable for a first
generation of our SETI efforts in the first decade, intended for urgent
probabilistic mitigation of short-term anthropogenic risks, while
a 100 ly range may be covered later to be conclusive with certainty
about absence of contactable ETI of this kind in our proximity. For
a similar reason, sensitivity of SETI can be improved in a second
generation if there are doubts about argumentation of N16 as minimal
beacon size.

\section{Beacon Schedule}

There must be slacks within the schedule of usual interstellar transmissions
where an interstellar transceiver may be used as a beacon without
sacrificing usual traffic, thus almost for free. Typical regular transmission
duration will be on the order of one month as suggested by orbital
motion of planets, with a typical slack on the order of one week.
Taking into account uncertainties, minimal duration of beacon transmission
may be estimated as 24 hours.

In an optimistic case, the period may be twice per year, but it may
be much less frequent, say once per decade for economic reasons, to
save resources of reaction wheels. But this is for the case of $S\geq50\%$.
If $S=1\%$, each planet may have 100 search targets; then the optimistic
case will be one decade and the pessimistic one century.

If existential risks of young civilizations on the verge of maturation
are objectively real and timely contact may help, ETI will care about
this to avoid failures of a young civilization to make contact, adjusting
the period of beacon signal accordingly, without resort to excessive
economy of reaction wheels. Thus, for the purpose of mitigation of
existential risks, the optimistic case can be assumed, though this
assumption has little practical sense. It only means that we have
a subjectively high (on the order of unity) but not quantifiable chance
to detect a signal within 10 yr. If not detected, we still must continue
indefinitely until we have no better idea. This uncertainty follows
from our own uncertainty about our own existential risks.

Without an urgent existential risk argument, ETI has no clear reasons
to support periods much shorter than the timescale of building potentially
destructive interstellar missions by a young civilization. Moreover,
ETI motives to support beacons can be marginal due to extreme rarity
of appropriate indigenous civilizations, but this will not lead to
not supporting beacons at all, because ETI always has the option to
increase the period to an extent where the expense of a beacon can
be supported just by inherent indispensable curiosity of any intelligent
beings.

It logically follows that the existence of anticipated contactable
ETI cannot be contradicted completely, even after 1000 yr of supporting
an appropriate contact strategy. Any time-limited support with a negative
result only updates a lower bound on the beacon period, but that bound
has high scientific value. Moreover, continuous support of a contact
strategy is more than just a scientific enterprise; it appears as
a mandatory prerequisite of a civilization intending to become a part
of a benevolent interstellar community.

\section{SETI Instruments}

\subsection{Required Size of Simplest SETI Instrument}

Here is the simplest model for a small telescope with a simple passband
optical filter corresponding to the optimal working band of a Yb:YAG
laser. Here the beacon is not resolved by the telescope spatially
against the host star, and the signal will be seen as a small bump
over a baseline dominated by starlight. Here sensitivity is limited
by shot noise, i.e., by random variation of photon number received
during exposure, whose natural Poissonian distribution is approximated
by a normal distribution.

The required diameter of a SETI telescope can be calculated as 
\begin{equation}
D_{s}=\sqrt{\frac{Bs}{\pi\eta_{s}t_{e}}}\frac{hc\lambda dq}{\eta_{t}P_{b}KN^{2}}=0.236\text{ m}
\end{equation}
where $d=50\text{ ly}$ is distance to target, $t_{e}=20\text{ hr}$
is exposure time, $s\approx10^{29}$ is spectral density parameter
for a Sun-like star at 1030 nm, $B=113\cdot10^{9}\text{ Hz}$ is spectral
band of the telescope (optimal band of Yb:YAG laser),$P_{b}=2000\text{ W}$
is beamline power, $\eta_{s}=\eta_{t}=0.2$ are photonic efficiencies
of the SETI instrument and transmitter respectively, $\lambda=1030\text{ nm}$
is wavelength, $K=10\text{ m}^{2}$ is beamline cost factor, $N=16$
is number of beamlines, $q=\mathrm{qnn}(P_{fns})+\mathrm{qnn}(P_{fp})=5$
is noise margin defined by given conditional false positive ($P_{fp}$)
and false negative ($P_{fns}$) probabilities, and $x(y)=\mathrm{qnn}(y)$
is the inverse of $y(x)=\frac{1}{2}\mathrm{erfc}\left(\frac{x}{\sqrt{2}}\right)$.
All equations in this article use the SI unit system if not stated
explicitly otherwise, even if numeric values are presented in other
units for convenience.

\subsection{Detectability by Known SETI Efforts}

Performance of a given SETI instrument can be expressed by the size
of a detectable beacon: 

\begin{equation}
N=\sqrt{\sqrt{\frac{Bs}{\pi\eta_{s}t_{e}}}\frac{hc\lambda dq}{\eta_{t}P_{b}KD_{s}}}=\sqrt{\sqrt{\frac{cbs}{\pi\eta_{s}t_{e}}}\frac{hcdq}{\eta_{t}P_{b}KD_{s}}}
\end{equation}
where $b$ is bandwidth in units of wavelength, such that $B=\frac{cb}{\lambda^{2}}$.

\subsubsection{All-Sky All-Time LaserSETI}

Ongoing LaserSETI efforts based on Gillum et al.\cite{Gillum (2022)}use
a small aperture ($D_{s}=0.017\text{ m}$) and low spectral resolution
system $\left(b=30\text{ nm}\right)$, and a silicon sensor which
can only be marginally sensitive at 1030 nm $\left(\eta_{s}=0.01\right)$.
Even if optics are assumed near diffraction-limited (which is not
the case because of extreme field of view), the minimal detectable
size of a beacon is $N=371$. Even if equipped with an appropriate
sensor $\left(\eta_{s}=0.2\right)$, it gives $N=175$.

\subsubsection{Targeted Spectroscopic Observations}

A marginal chance exists that a beacon can be discovered by targeted
astronomical spectroscopic observations of stars, when an exposure
occasionally overlaps with signal transmission in the time domain.

\section{Conclusions}

Existence of benevolent communicative ETI in our proximity is highly
probable and it is not contradicted by ``Fact A'' of Hart and the
fact of the ``Great Silence''. An appropriate strategy to contact
such ETI is shown to be affordable (below \$1/year per capita) and
probably can discover ETI within a decade. But there are no confident
bounds on time required for such discovery. Any continuous time-limited
support of the strategy without a positive outcome can only adjust
the lower bound on periodicity of the beacon. Knowledge of that bound
is of high scientific value because it is useful for statistical inference
of bounds on life in the Galaxy. Moreover, indefinitely continuous
support of this strategy is not just a scientific enterprise; it appears
as a mandatory prerequisite of a planetary civilization intending
to become part of a benevolent interstellar community, and as an effective
countermeasure against anthropogenic existential risks.

From the perspective of known SETI approaches, the novelty of the
proposed strategy is systematic use of principles of the top-down
approach, introduced here, which is useful in a quasimetastable state
only and therefore was not useful in the 20th century. A quasimetastable
state is defined here as an apparently mature state of technology,
which objectively may be either a mature or metastable state of technology.
Apparently, humanity crossed the verge of a quasimetastable state
at the beginning of the 21st century, and incapacity of known SETI
approaches can be explained by long-standing conservative adherence
to historical concepts invented in the 20th century without appropriate
critical analysis according to the current state of technology. Systematic
lack of such critical analysis can be explained by marginal motivation
of fundamental SETI research due to historically dominating pessimism.

Numeric results presented here are based on coarse calculations, simplified
models, and engineering intuition, and therefore must be considered
preliminary. Nonetheless, results appear sufficient as a preliminary
proof of feasibility for the new paradigm of SETI presented here,
which now requires deeper investigations.

\appendix

\section{Photonic Information Efficiency Approximation\label{sec:Appendix-A}}

Here a slotted OOK channel with single-photon counting (not photon-number
resolving, also known as SPC-OOK\cite{Erkmen (2012),Dolinar (2012)})
is assumed as an appropriate model for the purpose of this article
to get a coarse approximation for any practically optimal channel
where transmitted power is the dominant cost factor. The slotted SPC-OOK
model is already known as asymptotically approaching close to the
information-theoretical limit for power-limited channels, and this
is confirmed by a number of theoretical and experimental studies for
deep-space communications using already available technological goods\cite{Erkmen (2012),Dolinar (2012)}.
The analytical model of slotted SPC-OOK is a memoryless binary asymmetric
channel\cite{Moser(2010),Mathar(2018)} corresponding to one isolated
slot. This model assumes each real slot as an independent channel;
hence inter-slot leakage of photons in a real channel is neglected.
In the power-limited regime, the probability of the on-state per slot,
$p_{1}$, is low; hence $p_{1}\ll1$ can be assumed. Technically this
channel corresponds to low-duty pulsed transmission where pulse duration
fits within one slot time $t$ and average pulse frequency is $f=p_{1}/t$.
Then bitrate is $R=\frac{C}{t\ln(2)}$, where $C$ is capacity of
the memoryless binary channel of one slot. The nat is used here as
the unit of channel capacity and entropy, for convenience of mathematical
analysis, while bits are used for other quantities related to bitrate:
$1\text{ nat}=\frac{1}{\ln(2)}\text{ bits}\approx1.44\text{ bits}$.

According to information theory\cite{Mathar(2018)}, channel capacity
of a memoryless binary channel can be found as

\begin{equation}
C=\max_{p_{1}}I(X;Y)
\end{equation}
where $p_{1}$ is probability of the on-state of the input distribution
and 
\begin{equation}
I(X;Y)=H(Y)-H(Y|X)=\mathrm{h}\left((1-p_{1})(1-\epsilon_{0})+p_{1}\epsilon_{1}\right)-(1-p_{1})\mathrm{h}(\epsilon_{0})-p_{1}\mathrm{h}(\epsilon_{1})
\end{equation}
is mutual information, where:
\begin{description}
\item [{$\epsilon_{0}=\Pr(y=1\mid x=0)$}] is probability of detection
given off-state input, due to background photons;
\item [{$\epsilon_{1}=\Pr(y=0\mid x=1)$}] is probability of no detection
given on-state input, due to photonic loss;
\item [{$\mathrm{h}(x)=-x\ln(x)-(1-x)\ln(1-x)$}] is the binary entropy
function (nat convention).
\end{description}
Let $n_{s}$ be received pulse energy, measured in average photon
number, and $n_{b}$ be average number of received background photons
per slot. Then, assuming a Poissonian distribution of photon numbers
per slot and not resolving more than one photon detection per slot,
$\epsilon_{0}=1-e^{-n_{b}}$ and $\epsilon_{1}=e^{-n_{s}}e^{-n_{b}}$.
Received mean power, $p=p_{1}n_{s}$, is a given constraint for a
power-limited channel, where capacity is maximized for given $p$.
Hence $n_{s}$ becomes dependent on $p_{1}$, and it must be co-optimized
with $p_{1}$. Thus
\begin{equation}
I(X;Y)=\mathrm{h}\left(\left(1-\frac{p}{n_{s}}\right)e^{-n_{b}}+\frac{p}{n_{s}}e^{-n_{s}-n_{b}}\right)-\left(1-\frac{p}{n_{s}}\right)\mathrm{h}\left(e^{-n_{b}}\right)-\frac{p}{n_{s}}\mathrm{h}\left(e^{-n_{s}-n_{b}}\right)\label{eq:I2-1}
\end{equation}
\begin{equation}
C=\max_{n_{s}}I(X;Y)
\end{equation}

Let $s=e^{-n_{s}}$ and $b=e^{-n_{b}}$:

\begin{equation}
I(X;Y)=\mathrm{h}\left(\left(1+\frac{p(1-s)}{\ln(s)}\right)b\right)-\left(1+\frac{p}{\ln(s)}\right)\mathrm{h}(b)+\frac{p}{\ln(s)}\mathrm{h}(sb)
\end{equation}

\begin{equation}
\frac{d}{ds}I(X;Y)=\frac{-pb}{\ln(s)}\left[\left(1+\frac{1-s}{s\ln(s)}\right)\ln\left(\frac{1}{b\left(1+p\frac{1-s}{\ln(s)}\right)}-1\right)+\ln\left(\frac{1}{bs}-1\right)+\frac{\mathrm{h}(b)+\mathrm{h}(bs)}{bs\ln(s)}\right]
\end{equation}
from where the optimal value of $n_{s}$ can be found by solving $\frac{d}{ds}I(X;Y)=0$:

\begin{equation}
\left(1+\frac{1-s}{s\ln(s)}\right)\ln\left(\frac{1}{b\left(1+p\frac{1-s}{\ln(s)}\right)}-1\right)+\ln\left(\frac{1}{bs}-1\right)+\frac{\mathrm{h}(b)+\mathrm{h}(bs)}{bs\ln(s)}=0
\end{equation}
from where $p(n_{s},n_{b})$ can be exactly calculated as

\begin{equation}
p(n_{s},n_{b})=\begin{cases}
s:=\exp(-n_{s})\\
b:=\exp(-n_{b})\\
z:=\frac{1-s}{s\ln(s)}\\
q:=b\left(1+\exp\left(-\frac{\ln\left(\frac{1}{bs}-1\right)+\frac{\mathrm{h}(b)+\mathrm{h}(bs)}{bs\ln(s)}}{1+z}\right)\right)\\
\frac{\frac{1}{q}-1}{zs}
\end{cases}
\end{equation}
and then a closed-form approximation for $n_{s}(p,n_{b})$ can be
expressed as
\begin{equation}
n_{s}(p,n_{b})=\begin{cases}
f(p):=\frac{2.18}{\ln(\ln(32/p))}-0.375+\frac{p^{2}}{4}\\
m:=-3\ln(n_{b})\\
\frac{\ln\left(e^{mf(p)}+e^{mf(2n_{b})}\right)}{m}+(1.5+6p)n_{b}
\end{cases}\label{eq:ns}
\end{equation}
accurate for $n_{b}<0.1$ and $p<3$.

Then Photonic Information Efficiency (PIE), in bits per photon, can
be calculated as
\begin{equation}
\Xi(p,n_{b})=\frac{C}{p\ln(2)}=\begin{cases}
n_{s}:=n_{s}(p,n_{b})\\
C:=\mathrm{h}\left(\left(1-\frac{p}{n_{s}}\right)e^{-n_{b}}+\frac{p}{n_{s}}e^{-n_{s}-n_{b}}\right)-\left(1-\frac{p}{n_{s}}\right)\mathrm{h}\left(e^{-n_{b}}\right)-\frac{p}{n_{s}}\mathrm{h}\left(e^{-n_{s}-n_{b}}\right)\\
\frac{C}{p\ln(2)}
\end{cases}
\end{equation}
and then the simple approximation

\begin{equation}
\Xi=1.22\left(-\ln(3n_{b}+p)-1\right)
\end{equation}
is useful for $n_{b}<0.01$ and $p<0.01$.

Recall $p$ is average received signal photons per slot, $n_{b}$
is average received noise photons per slot, $I(p,n_{b})$ is average
nats per slot, and the model used is SPC-OOK, which gives an approximate
capacity limit of a deep-space power-limited channel for $p\ll1$
and $n_{b}\ll1$. The rationale of this last heuristic approximation
can be viewed as ignoring irrelevant behavior of the model beyond
its usability range. Thus, accounting for known inherent limitations
of the model, this approximation is not necessarily less accurate
than the exact solution based on the SPC-OOK model.

\end{document}